\documentclass[12pt]{article}
\usepackage{fullpage}
\usepackage{amssymb, amsthm, amsmath}
\usepackage{doublespace}
\usepackage{graphicx}
\usepackage[authoryear]{natbib}
\usepackage{bm}
\usepackage{verbatim}
\usepackage{lineno}
\usepackage{times}
\usepackage{tabulary}
\usepackage[hyphens]{url}
\usepackage{hyperref}
\usepackage{breakurl}
\usepackage{gensymb}
\usepackage{subfig}
\usepackage{arydshln}
\usepackage{hyperref}

\newcommand{\bbeta}{ \mbox{\boldmath $\beta$}}

\newcommand{\bpsi}{ \mbox{\boldmath $\psi$}}

\newcommand{\bSigma}{ \mbox{\boldmath $\Sigma$}}

\newcommand{\be}{ \mbox{\bf e}}

\newcommand{\bx}{ \mbox{\bf x}}
\newcommand{\bX}{ \mbox{\bf X}}

\newcommand{\by}{ \mbox{\bf y}}
\newcommand{\bY}{ \mbox{\bf Y}}

\newcommand{\bs}{ \mbox{\bf s}}

\newcommand{\bv}{ \mbox{\bf v}}
\newcommand{\bV}{ \mbox{\bf V}}

\newcommand{\bfe}{ \mbox{\bf e}}
\newcommand{\bzero}{ \mbox{\bf 0}}
\newcommand{\iid}{\stackrel{iid}{\sim}}
\newcommand{\indep}{\stackrel{indep}{\sim}}

\newcommand{\calL}{{\cal L}}

\newcommand{\calC}{{\cal C}}

\newcommand{\beq}{ \begin{equation}}
\newcommand{\eeq}{ \end{equation}}
\newcommand{\beqn}{ \begin{eqnarray}}
\newcommand{\eeqn}{ \end{eqnarray}}

\begin{document}

\begin{center}
{\Large Fine-scale spatiotemporal air pollution analysis using mobile monitors on Google Street View vehicles}\\\vspace{6pt}
{\large \footnote{The first two authors contributed equally to this work}Yawen Guan\footnote{North Carolina State University}\footnote{The Statistical and Applied Mathematical Sciences Institute}, $^1$Margaret C Johnson$^{23}$, 
Matthias Katzfuss\footnote{Texas A\& M University},
Elizabeth Mannshardt\footnote{US Environmental Protection Agency}, Kyle P Messier\footnote{Oregon State University}, Brian J Reich$^2$ and
Joon Jin Song\footnote{Baylor University}
}\\
\today
\end{center}
\begin{abstract}\begin{singlespace}
\noindent People are increasingly concerned with understanding their personal environment, including possible exposure to harmful air pollutants. In order to make informed decisions on their day-to-day activities, they are interested in real-time information on a localized scale. Publicly available, fine-scale, high-quality air pollution measurements acquired using mobile monitors represent a paradigm shift in measurement technologies. A methodological framework utilizing these increasingly fine-scale measurements to provide real-time air pollution maps and short-term air quality forecasts on a fine-resolution spatial scale could prove to be instrumental in increasing public awareness and understanding.  The Google Street View study provides a unique source of data with spatial and temporal complexities, with the potential to provide information about commuter exposure and hot spots within city streets with high traffic. We develop a computationally efficient spatiotemporal model for these data and use the model to make short-term forecasts and high-resolution maps of current air pollution levels.  We also show via an experiment that mobile networks can provide more nuanced information than an equally-sized fixed-location network.  This modeling framework has important real-world implications in understanding citizens' personal environments, as data production and real-time availability continue to be driven by the ongoing development and improvement of mobile measurement technologies.\vspace{12pt}\\

\end{singlespace}\end{abstract}
\newpage

\section{Introduction}\label{s:intro}


The harmful effects of air pollution on human health have been well documented \citep{pope1995review, pope2006health, chang2011time, laden2006reduction, zanobetti2009effect, katsouyanni2009air}. The World Health Organization classified air pollution as a major environmental health risk, estimating that 4.2 million premature deaths in 2016 can be attributed to exposure to outdoor air pollution \citep{world2016ambient}. Nitrogen dioxide (NO$_2$), a highly reactive gas monitored by states and the EPA, is formed primarily from fuel-burning emissions.  Health effects at elevated levels from short-term exposures include cardiovascular effects and premature mortality as well as difficulty breathing and increased occurrence of hospital visits due to decreased lung capacity including asthma exacerbation \citep{epa2016nitrogen}.  Long-term effects at elevated levels include cardiovascular effects, premature mortality, diabetes, poorer birth outcomes, cancer, and asthma in children \citep{epa2016nitrogen}.

NO$_2$ also reacts with water in the atmosphere to produce ozone and acid rain.  The resulting nitrate particles from this reaction can contribute to particulate matter less than 2.5 microns in diameter (PM$_{2.5}$) \citep{epa1999nitrogen}, all of which have additional environmental health considerations. NO$_2$ can react with water, ozone, and nitric oxide (NO) multiple times over a span of several hours to form and re-form NO$_2$ and NO.  Emissions of nitrous oxides are primarily in the form of NO, where 92\% of NO is anthropogenic with 56\% estimated to be from mobile emissions \citep{epa2014nei}. 
Hence understanding patterns of mobile emissions are paramount to protecting public health, particularly for susceptible populations, such as children, the elderly, and those with asthma or compromised immune systems.

The vast majority of health studies to date have focused on the relationship between human health effects and longer term exposures, such as 1-hour or daily average aggregate effects.  Additionally, these studies are based on air-quality measurements for which spatial information is often limited due to the number of stationary monitors measuring air quality over large regions.  It is important to understand these pollutant patterns on a finer scale, as microenvironment effects can vary strongly according to meteorology and local traffic patterns. Very fine-scale measurements are becoming more relevant as portable and mobile sensors are deployed by local governments in an effort to assess, evaluate and manage local air quality conditions. This includes Chicago's ``Array of Things" citywide network of air quality sensors and Louisville's Air Louisville initiative launched in 2012 to monitor asthma-enducing conditions \citep{aq2015threecities}. International efforts include Air Map Korea Project \citep{aq2018iot}, which aims to install over 4.5 million monitors on telephone poles, public phone booths and central offices, and Smart City Barcelona's Lighting Masterplan \citep{aq2016barcelona}, which equipped lampposts with air quality sensors to relay information to the city and to the public. London showed creativity in air quality monitoring with their Pigeon Patrol \citep{aq2016pigeons}, fitting pigeons with mobile air quality sensors to measure nitrogen dioxide across the city.  Communication of these short-term measurements is a significant challenge.  Methodology utilizing this information to provide real-time air pollution maps as well as short-term air quality forecasts on a fine-resolution temporal and spatial scale may revolutionize people's understanding of their personal environment and exposures, having real-world implications and impacts on citizens.  

As the effects of air pollution and the differences in pollutant microenvironments become more widely studied and understood, people are increasingly concerned with understanding their immediate personal environment and its effect on their health.  They are interested in real-time information on a very localized scale in order to make informed decisions on their day-to-day activities.  This includes their possible exposure to harmful air pollution. They may routinely consider questions such as: What is the best time of day to go for a run through a residential neighborhood?  What route should an asthmatic take to work to avoid high levels of air pollution?  Are air pollutant levels at a city park higher or lower during an afternoon or evening weekday versus a weekend? EPA's widely used Air Quality Index (AQI) uses high quality hourly data from stationary Federal Reference Monitors (FRMs) and Federal Equivalency Monitors (FEM) to implement a color-coded air quality scale and public messaging system.  Real-time and forecasted AQI's are provided to the public via EPA's AirNow website (\url{airnow.gov}) and are available by city, state, or ZIP code. Localized sensor networks, which may include mobile sensors, may provide additional information on which to build, refine, and expand air quality information available to the public.  Additionally, as technology continues to improve, personal wearables are becoming more affordable and accessible to the general public \citep{aq2017flow,aq2017plume}. 

Real-time information on a very localized scale are also valuable for being able to consider possible health effects of very short-term exposures.  It is important to first be able to describe the behavior of pollutants on the corresponding spatial and temporal scales in order to consider possible health effects at such a fine scale.

In this paper, we analyze data collected from Google Street View vehicles in Oakland, CA \citep{apte2017high}.  The cars drive through the city on spatiotemporal tracks and measure ambient NO$_2$ each second.  This study provides a unique source of highly detailed data with spatial and temporal complexities.  It can provide information about commuter exposure, hot spots within high-trafficked city streets, as well as complex patterns due to meteorological effects and microenvironments.  This fine-scale spatial and temporal information could also lead to the methodology and information needed to start to characterize acute exposure.  It is particularly important to understand near-road and city-street environments.  The US Environmental Protection Agency reports that over 45 million people live in close proximity to major roadway \citep{epa2018nearroad}.  The novel data set provides information about air quality surrounding roadways and commonly trafficked areas that is not available from the limited number of stationary monitors across an area or region.
 

Fine-scale air quality measurement and analysis have been powered by recent advances in sensor technologies, which allow for the use of mobile network platform with low-cost sensors for the purpose of general monitoring {\citep{snyder2013changing,morawska2013indoor,CASTELLINI2014,Sarto2016}} and personal exposure assessment \citep{holstius2014field,kim2012automatic,white2012sensors}. There is a significant difference in temporal resolution between mobile measurements and fixed point networks.  As a complement to fixed point networks, mobile air quality monitoring can also improve spatial coverage and be used to map air pollution with improved spatial and temporal resolution. In particular, fine-scale air quality monitoring is essential in urban settings, because the measurements vary dramatically over space and time and closely relate to several factors, such as land use, traffic, and meteorology. Several mobile platform have been developed, such as wearable device (Hu et al., 2014), smart-phone \citep{dutta2009common,hasenfratz2012participatory,bartonova2015citi,brienza2015low}, bicycle \citep{thai2008particulate,boogaard2009exposure} and vehicle \citep{gulliver2007journey, larson2007spatial,briggs2008effects,boogaard2009exposure}. Recently, \cite{apte2017high} demonstrated fine-scale spatial (though not spatiotemporal) air pollution mapping with the largest urban air quality data collected by Google Street View vehicles.

In this paper, we perform a novel spatiotemporal analysis of Google Street View data.  Our objectives are to develop a statistical approach to use these data to produce fine-scale maps of the current air pollution level and to make short-term local forecasts.  These data are streaming (i.e., collected every second) and collected along spatiotemporal paths (cars traversing the city), providing new insights into air quality via a mobile measurement framework.  We tailor our computational approach to these unique features using local-likelihood approximations. We explore different forms of temporal aggregation to compare stability by aggregation level and address the practical problem of selecting the neighborhood scheme for the local approximation to optimize short-term prediction.  We show that our final approach has forecast skill and outperforms competing methods.  Finally, we conduct a simulated experiment to determine the relative effectiveness of a fleet of mobile monitors compared to a network of fixed-location monitors.  We find that mobile monitors provide improved estimation and prediction at un-monitored locations compared to estimation and prediction based solely on limited fixed-location monitors.  Therefore, our paper contributes to the emerging field of mobile air pollution monitoring by providing a template for processing and modeling data with these types of complex measurement and scale considerations, as well as guidance for future sampling efforts.

\section{Data sources and exploratory analysis}\label{s:data}

One-second NO$_2$ data were collected via routine mobile monitoring in Oakland, California as part of an on-going multi-institutional collaboration between the Environmental Defense Fund, University of Texas at Austin, and Google, among others. Details of the sampling protocol are available in \cite{apte2017high}. Briefly, the data were collected with two Google Street View mapping vehicles (henceforth referred to as Car A and Car B) equipped with Aclima-Ei fast-response pollution data integration platform (See \cite{apte2017high} Supporting Information for details; Aclima Inc., San Francisco CA). {The measurement instrument is mounted on the top of the vehicle to alleviate the effect of high emission due to self-emission from the vehicles. Post-installation tests indicated self-sampling was a rare occurrence during routine operation, including idles and stops \citep{apte2017high}.} Data were collected during weekdays and daytimes, in which the drivers were instructed to drive every road segment at least once in an assigned polygon with area between 1-10 km$^2$. In this study, pollutant data were obtained directly and used with permission from Google. Data included samples from June 30, 2015 to May 13, 2016. 
Following \cite{apte2017high}, we employed the following data reduction algorithm to convert raw spatial coordinates into consistently defined spatial locations. First, a street centerline file (obtained from \url{OpenStreetMaps.com}) was converted to roughly equal interval 30-meter road segments. Next, a nearest-neighbor algorithm assigned the raw geographical coordinates to the nearest 30-meter segment resulting in consistently defined locations to evaluate spatial and temporal trends. 

For each segment we also extracted the 28 geographic covariates.  We include binary indicators for non-residential road types (highway, major road, truck route, and major truck route) and land-use zones (commercial, industrial and residential).  Continuous variables include distance to point sources (railway, port, airport, EPA superfund National Priority Listing sites and EPA Toxic Release Inventory sites) and the average value of several variables with in a 50m buffer around the segment: elevation, population, normalized difference vegetation index to measure greenness, several land cover types (water, open developed, low developed, medium developed, high developed, evergreen, shrub, herbaceous and impervious) and lengths of types of roads (total, highway, major and residential).

\begin{figure}
 \centering
 \includegraphics[width=\textwidth]{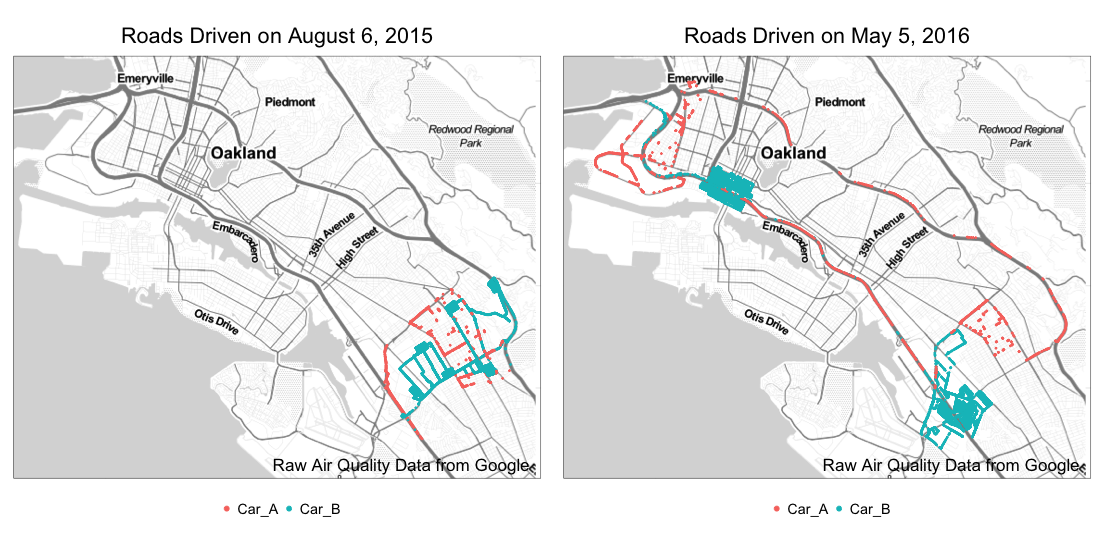}
 \caption{{\bf Example Drive Coverage:} Driving routes of two Google Street View vehicles for two days. The gray lines represent all the road segments covered by the two vehicles within the data collection period. Map tiles by \href{http://stamen.com}{Stamen Design}, under \href{http://creativecommons.org/licenses/by/3.0}{CC BY 3.0}. Data by \href{http://openstreetmap.org}{OpenStreetMap}, under \href{http://www.openstreetmap.org/copyright}{ODbL}.}
 \label{fig:routes}
 \end{figure}

Car routes driven on August 6th, 2015 and May 5th, 2016 for the two Google Street View vehicles are shown in Figure \ref{fig:routes}. On August 6th, both cars drove mainly in the residential areas of the southeast region of Oakland between 9:00 and 15:00. On May 5th, Car A drove mainly highways throughout Oakland while Car B covered two mostly residential areas, both cars driving between 10:00 and 15:00. The two days illustrate that while air pollution data can be obtained from the Google Street View vehicles locally at very high spatiotemporal resolution, the overall space-time coverage of the data on a given day is quite limited. In addition, both cars were driven simultaneously on only 41\% of the drive days, with only one car driving on the remaining days.  It is also important to note NO$_2$'s diurnal pattern, with elevated levels due to emissions in the later morning (\cite{pancholi2018}).

\begin{figure}[ht]
 \centering
 \includegraphics[width=\textwidth]{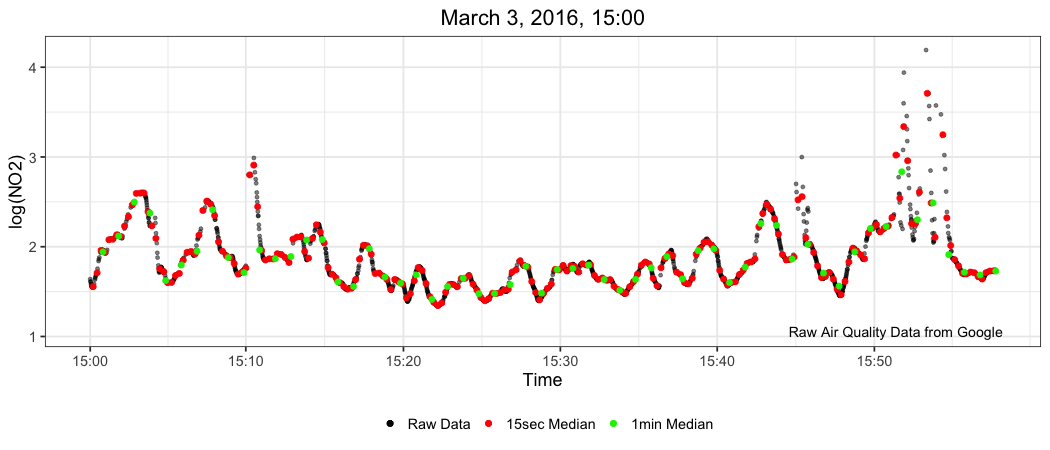}
 \caption{{\bf Temporal Aggregation:} Illustration of data reduction via block medians for one hour of data.}
 \label{fig:blocks}
 \end{figure}

Exploration of the one-second samples shows that the NO$_2$ data are heavily right-skewed, even after a log transformation.  We hypothesize that the extremely large values are caused by very local and unpredictable phenomena such as the car being stuck behind a heavy-polluting vehicle.  Therefore, in addition to modeling the one-second data we also consider using medians of the samples over 15-second and one-minute intervals {to lessen the influence of these extreme values}. Figure \ref{fig:blocks} plots a sample of the data along with corresponding block medians.  The spatial location assigned to these block medians is the location of the sample nearest to the center of the block's time interval. Therefore, a trade-off for using the more stable block median data is loss of spatial fidelity.
 
\section{Statistical model}\label{s:model}

Let $Y_t(\bs)$ be the observed log-transformed NO$_2$ from a car measured at time $t$ and location $\bs \in \mathbb{R}^2$. $Y_t(\bs)$ is regressed onto $\bX_t(\bs)$, a $p$-dimensional vector of covariates at location $\bs$ and time $t$. We assume the observed log-transformed NO$_2$ is a noisy realization of the true NO$_2$ process, $\mu_t(\bs)$, such that $Y_t(\bs)|\mu_t(\bs)\indep\mbox{Normal}(\mu_t(\bs),\tau^2)$.  The true process $\mu_t(\bs)$ is decomposed using the spatiotemporal land-use regression model
\beq\label{e:DLM}
\mu_t(\bs) = \bX_t(\bs)^T\bbeta + \eta_t(\bs).
\eeq
The model allows $\bX_t(\bs)^T\bbeta$ to capture the large-scale spatiotemporal variations in NO$_2$ while the remaining variability is captured by the small-scale process $\eta_t(\bs)$. 

The small-scale process follows a Gaussian process with mean zero and spatiotemporal covariance function $ \mathcal{C}(\bs,t,\bs',t')=\mbox{Cov}[\eta_t(\bs),\eta_{t'}(\bs')]$, denoted $\eta\sim\mbox{GP}(0,\calC)$. We assume a nonseparable, nonstationary covariance function where distance is a function of space, time, and (standardized) spatial covariates $\bx(\bs)$ (that need not be the same covariates as in $\bX_t(\bs)$), 
\beq\label{e:C}
	\mathcal{C}(\bs,t,\bs',t') = 
    \sigma^2 \text{exp}\left(-\sqrt{\dfrac{||\bs - \bs'||^2}{\theta_s^2} + \dfrac{(t - t')^2}{\theta_t^2} + \dfrac{||\bx(\bs) - \bx(\bs')||^2}{\theta_x^2}}\right). 
\eeq 
The covariance has four parameters: $\sigma^2$ is the variance, and $\theta_s$, $\theta_t$ and $\theta_x$ control the range of correlation in space, time and covariate space, respectively. The covariance is an anisotropic exponential covariance function defined using Euclidean distance on the $(p+3)$-dimensional domain spanned by $[\bs,t,x(\bs)]$.

The covariance is a function of both spatial distance $||\bs-\bs'||$ and similarity between the covariates, $\bx(\bs)$, at sites $\bs$ and $\bs'$, as well as how close observations are in time.  Including covariates in covariance \citep[e.g.,][]{schmidt2011considering,reich2011class,ingebrigtsen2014spatial,risser2015regression} allows for the model to extrapolate to areas far from the observations but with similar geographic profiles.  For example, it may be that all locations near a highway are high due to heavy traffic that day, and using distance to highway as a covariate in the covariance captures this complex dependence structure.  Allowing for this rich dependence structure is potentially useful for the Google Street View application where the spatiotemporal coverage is often sparse.

\section{Computational details}\label{s:comp}

The Google Street View data are large and we {intend to refit the model periodically to update parameter estimates using the most recent data}, therefore a fully Bayesian or maximum likelihood analysis is infeasible. In this section we describe a computationally efficient approximation {of the likelihood function} that can be applied to large spatiotemporal data. {A two-step procedure is adapted for parameter estimation.} We first estimate the regression coefficients, $\bm{\beta}$, from a multiple linear regression model assuming independent errors. The residuals $\hat{\epsilon}_t(\bs)=Y_t(\bs)-\bX_t(\bs){\hat \bbeta}$ are then used to estimate the remaining parameters $\Theta = (\tau^2,\sigma^2, \theta_t, \theta_s, \theta_x)$ using composite maximum likelihood estimation.

The $n$ residuals are ordered in time with ${\hat \epsilon}_i$ denoting the $i^{th}$ residual for the observation at location $\bs_i$ and time $t_i$, i.e., ${\hat \epsilon}_i = {\hat \epsilon}_{t_i}(\bs_i)$.
We use the Vecchia composite likelihood \citep{vecchia1988estimation,stein2004approximating,katzfuss2017general,guinness2018permutation}  
\begin{equation}\label{e:cmle}
  \calL(\Theta) = \prod_{i=1}^nf\left({\hat \epsilon}_i|{\hat \bfe}_i,\Theta\right), 
\end{equation}
where ${\hat \bfe}_i$ is the vector of ${\hat \epsilon}_j$ for $j\in{\cal N}_i$, ${\cal N}_i\subseteq\{1,...,i-1\}$ is the conditioning set for observation $i$, and $f$ is the conditional distribution of ${\hat \epsilon}_i$ given ${\hat \bfe}_i$.  Specifically, $f({\hat \epsilon}_i|{\hat \bfe}_i,\Theta)$ is the Gaussian density function with mean $\bSigma_{i12}(\Theta) \bSigma_{i22}(\Theta)^{-1}{\hat \bfe}_i$
and variance $\bSigma_{i11}(\Theta) - \bSigma_{i12}(\Theta) \bSigma_{i22}(\Theta)^{-1} \bSigma_{i12}(\Theta)^T$, where $\bSigma_{i11}(\Theta)=\text{Var}({\hat \epsilon}_i)$, $\bSigma_{i12}(\Theta)=\text{Cov}({\hat \epsilon}_i,{\hat \bfe}_i)$ and $\bSigma_{i22}(\Theta)=\text{Cov}({\hat \bfe}_i)$ are determined by the spatiotemporal covariance function ${\cal C}$ with parameters $\Theta$.  This approximation is computationally convenient, because the largest matrix ($\bSigma_{i22}$) is the dimension of the conditioning set ${\cal N}_i$, which is taken to be much smaller than $n$. {In addition, the conditional distributions in the product can be computed in parallel using multiple processors. Our current implementation utilizes the \textit{parallel} \citep{parallel} R package for parallel computing.}

The key to effectively applying the Vecchia approximation is to select appropriate conditioning sets. If the conditioning sets include all previous observations, ${\cal N}_i = \{1,...,i-1\}$, then the composite likelihood is the exact likelihood.  While this may be optimal from the statistical perspective, it is infeasible from the computational perspective.  The composite likelihood is a close approximation to the full likelihood if the observations in the conditioning set explain most of the variability in the observations, and so a common approach is to take the conditioning set to be the closest neighbors to observation $i$.   However, recent work \citep{gramacy2015local} suggests that other conditioning sets that include distant points may lead to more precise estimation and prediction. 

We consider conditioning sets of the form
\begin{equation}\label{e:N}
  {\cal N}_i = \{j: t_i-t_j\in(l,l+m)\},
\end{equation}
i.e., the set of observations taken between $l$ and $l+m$ minutes before observation $i$. A natural choice is to set $l=0$ so that the conditioning set includes the most recent observations.  Given that the Google Street View data are collected along spatiotemporal paths, this conditioning set will include mostly observations taken close to $\bs_i$, and conditioning on these close spatiotemporal neighbors will provide a good approximation to the full likelihood.   

While setting $l=0$ provides a better approximation to the full likelihood, we find that conditioning sets with $l>0$ lead to better temporal predictions.  We hypothesize that this is because when $l=0$ the conditional distributions are dominated by a few neighbors that are very close to observation $i$, and when forecasting for moderate or large time lags no such neighbors are available.  Of course, if the parametric form of the covariance function is correct, then the covariance-parameter estimates derived from close neighbors can still give good distant predictions.  However, if $l=0$ and the parametric form of the covariance function is misspecified, then covariance estimates for long distances are extrapolations based on parameters estimated from short distances, and may be suboptimal.  Therefore, taking $l>0$ and thus excluding the closest points from the conditioning sets may give better estimates of the covariance function at the range needed for prediction; see Appendix A.1 for a small simulation study exploring this hypothesis for time series data.  

\section{Analysis of Oakland Google Street View data}\label{s:results}
We treat block medians with different time blocks as separate data sets and analyze them in parallel. {For our study period from June 30, 2015 to May 13, 2016, there are 901,215 measurements collected by two vehicles. The sizes of data sets with different time blocks are different; there are 132,347 block medians based on 15-second intervals, and 39,113 block medians based on 1-minute intervals.} We first conduct extensive model comparisons using data from 10/29/2015 to 12/18/2015 for training and the data from 12/21/2015 to 02/05/2016 for testing in Section \ref{s:compare}. Considering potential changes in the relationship between the land-use covariates and NO$_2$ and spatiotemporal dependence, due to evolving environmental, traffic and emissions patterns, we present the results of refitting the model periodically using a sliding window of training data in Section \ref{s:window}.  Finally, in Section \ref{s:experiment} we compare the efficiency of mobile versus fixed-location monitoring networks.

\subsection{Results for 10/29/2015 to 02/05/2016}\label{s:compare}

\begin{figure}
\centering
\includegraphics[width=\textwidth]{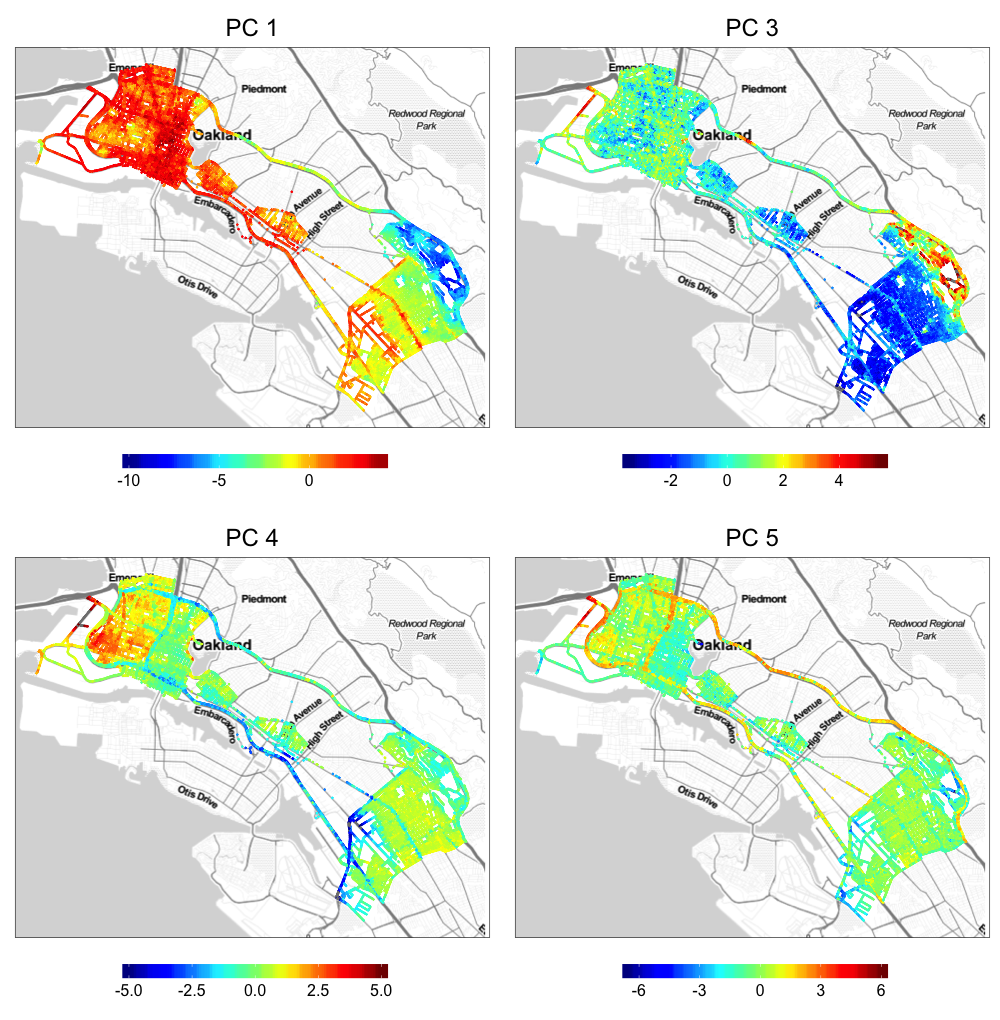}
\caption{{\bf Principal Components:} Spatial maps of principal components 1,3,4 and 5. Note that the color scale for each plot is different due to the large variation of values across PCs. Map tiles by \href{http://stamen.com}{Stamen Design}, under \href{http://creativecommons.org/licenses/by/3.0}{CC BY 3.0}. Data by \href{http://openstreetmap.org}{OpenStreetMap}, under \href{http://www.openstreetmap.org/copyright}{ODbL}.}
\label{f:pcs}
\end{figure}

Here we compare land-use regression models with different dependence structures. The mean components in the regression models use the same covariates derived from the geographic covariates given in Section \ref{s:data}. Because some of them are highly correlated, we standardize the covariates and compute their principal components (PC) to alleviate collinearity concerns. We then select a subset of these PCs in our analysis to reduce the dimension of the covariates. The first 13 PCs account for approximately 80\% of the variation in land-use covariates, and they capture important spatial features of the geographic covariates. As an example, four PCs are presented in Figure \ref{f:pcs}, the first PC is high for the downtown area and the fifth PC shows the contrast for major/non-major freeways.  To further reduce the number of covariates, we perform a least-squares analysis by regressing the 13 PCs onto log NO$_2$ and retain only the first seven PCs.
These seven PCs comprise the covariance covariate vector, $\bx(\bs)$.  In the mean term $\bX_t(\bs)$, we use the selected PCs, four diurnal covariates (cosine and sine terms with periods 12 and 24 hours) and their interactions, giving $p=40$ spatiotemporal covariates including the intercept. This allows the covariate effects to vary within day, for example, highway may have higher NO$_2$ concentration in the morning and evening due to rush hour traffic. 

The following models are compared:
\begin{enumerate}
  \item {\bf X-only}: The non-spatial land-use regression model $Y_t(\bs)\indep\mbox{Normal}\left(\bX_t(\bs)^T\bbeta,\tau^2\right)$.
  \item {\bf S}: The spatial land-use regression model $Y_t(\bs)|\eta(\bs)\indep\mbox{Normal}(\bX_t(\bs)^T\bbeta + \eta(\bs),\tau^2)$ where $\eta\sim \mathcal{GP}(0,\calC)$ with
  \beq
  \calC(\bs_i, \bs_j) = \sigma^2 \text{exp}\left\{ -\dfrac{||\bs_i - \bs_j||}{\theta_s} \right\}.
  \eeq
  \item {\bf ST}: The spatiotemporal land-use regression model $Y_t(\bs)|\eta_t(\bs)\indep\mbox{Normal}(\bX_t(\bs)^T\bbeta + \eta_t(\bs),\tau^2)$ where $\eta_t(\bs) \sim \mathcal{GP}(0, \calC)$ with 
  \beq 
  	\calC(\bs_i, t_i, \bs_j, t_j) = \sigma^2 \text{exp} \left\{ -\sqrt{\dfrac{||\bs_i -\bs_j||^2}{\theta_s^2} + \frac{(t_i - t_j)^2}{\theta_t^2}}\right\}.
  \eeq
  \item {\bf STx}: The full model with $Y_t(\bs)|\eta_t(\bs)\indep\mbox{Normal}(\bX_t(\bs)^T\bbeta + \eta_t(\bs),\tau^2)$, where $\eta_t(\bs) \sim \mathcal{GP}(0, \calC)$ with
  \beq 
  	\calC(\bs_i, t_i,\bs_j, t_j) = \sigma^2 \text{exp}\left\{-\sqrt{\dfrac{||\bs_i - \bs_j||^2}{\theta_s^2} + \dfrac{(t_i - t_j)^2}{\theta_t^2} + \dfrac{||\bx(\bs_i) - \bx(\bs_j)||^2}{\theta_x^2}}\right\}. 
 \eeq
\end{enumerate}
The first two models are static over days (although their means vary by hour of the day) to represent the spatial-only modeling approach of \cite{messier2018}.  Predictions from the final two models change by day based on data collected just prior to time the prediction is made.

Data from both cars in the training period are ordered in time for parameter estimation. The spatial-only model (S) is fitted using the Vecchia approximation with $30$ nearest neighbors in time in the conditioning set. 
For the ST and STx models, we repeat the proposed two-step procedure for different conditioning sets, i.e., different time lags ($l$) and neighborhood sizes ($m$). For one-second data, the conditioning sets are large even for small $m$. For instance, when $m=10$ minutes the conditioning set can have up to $600$ observations. To reduce computation time, we reduce the size of the conditioning set by subsampling $100$ observations for computing the composite likelihood. The results are similar for values of $m$ from 10 to 60 minutes for all different block medians, therefore, we present results only for $m=60$ minutes. In this case, the number of observations in the conditioning set is about $60$ for one-minute block median, which is an appropriate size for computing the composite likelihood.  For all methods, the parameters are estimated using only training data and never updated using test-set data.  For the spatial-only model we make Kriging predictions based only on the 800 nearest spatial neighbors in the training set. For the spatiotemporal methods (ST and STx) we make Kriging predictions based on the observations from the test-set data $l+m$ minutes prior to the time of the prediction.   

We perform two types of cross validation predictions on the test set: (1) forecast for $h=\{5, 15, 60\}$ minutes ahead given the past one hour observations and (2) predict one car conditioning on the data from the other car on the given day (denoted ``Car AB'' prediction). Forecast performance relies more on temporal dependence, because in our data the  cars cover only a small region in any given day, and so the forecast ability is limited to a small region near the latest available data. In Car AB prediction, the spatial dependence plays a larger role because the cars are often in different parts of the city.

\begin{table}
\caption{\textbf{Cross validation results.} Root mean squared prediction error (ppb) and prediction correlation for $h$ minutes ahead prediction and Car AB prediction using the mean-only (``X''), spatial-only (``S''),  spatiotemporal (``ST'') and covariates in covariance (``STx'') model.  The spatiotemporal models are fit using the observations from the previous $[l,l+60]$ minutes in the conditioning set.  The STx model is fit with $l=60$ for Car AB prediction.  The results are presented separately for data representing medians over one-second, fifteen-second, and one-minute block medians. }
\label{tab:CV}

\begin{center}\begin{tabular}{ccc|rr|rr|rr|rr}
Block &&& \multicolumn{2}{c}{$h=5$ min} & \multicolumn{2}{c}{$h=15$ min} & \multicolumn{2}{c}{$h=60$ min} &
\multicolumn{2}{c}{Car AB}\\
size & Model & $l$ & RMSPE & Cor & RMSPE & Cor & RMSPE & Cor & RMSPE & Cor\\\hline

&&&&&&&&&&\vspace{-6pt}\\ 
1 sec &X   &  -  &  10.79 & 0.18 & 10.79 & 0.18 & 10.79 & 0.18 &  55.29 &  0.08 \\ 
&S   &  -  &  11.22 & 0.27 & 11.22 & 0.27 & 11.22 & 0.27    &  109.80 &  0.09      \\ 
&ST  &  0  &  9.06 & 0.45 & 10.41 & 0.25 & 10.79 &  0.18    &  55.03 &  0.09  \\ 
&ST  &  5  &  8.22 & 0.58 & 10.98 & 0.36 & 10.78 & 0.28   &  55.01 &  0.10  \\ 
&ST  &  15 &  8.24 & 0.57 & 10.74 & 0.36 & 10.39 & 0.31   &  55.09 &  0.10 \\ 
&ST  &  60 &  8.47 & 0.55 & 9.99 & 0.38 & 10.33 & 0.28   &  55.00 &  0.09 \\ 
&STX &  h  &   8.26 & 0.58 & 10.46 & 0.37 & 10.71 & 0.25     &  55.12 & 0.10 \\ 
&&&&&&&&&&\vspace{-6pt}\\ 
15 sec &X   &  -  &  10.81 & 0.24 & 10.81 & 0.24 & 10.81 & 0.24 &  30.10 &  0.14 \\ 
&S   &  -  &  11.49 & 0.28 & 11.49 & 0.28 & 11.49 & 0.28    &  46.85 & 0.14      \\ 
&ST  &  0  &  9.90 & 0.44 & 10.68 & 0.28 & 10.81 & 0.24    &  29.94 &  0.17  \\ 
&ST  &  5  &  9.04 & 0.57 & 9.75 & 0.48 & 10.25 & 0.40   &  30.18 &  0.19  \\ 
&ST  &  15 &  9.04 & 0.57 & 9.75 & 0.48 & 10.28 & 0.40   &  30.16 &  0.19 \\ 
&ST  &  60 &  9.06 & 0.56 & 9.65 & 0.48 & 10.23 & 0.38   &  30.03 &  0.17 \\ 
&STX &  h  &  9.09 & 0.56 & 9.67 & 0.48 & 13.01 & 0.11     &  30.02 &  0.19 \\ 
&&&&&&&&&&\vspace{-6pt}\\      
1 min &X   &  -  &  9.61 & 0.28 & 9.61 & 0.28 & 9.61 & 0.28    &  20.99 &  0.19 \\ 
&S   &  -  &  10.36 & 0.34 & 10.36 & 0.34 & 10.36 & 0.34 &  23.40 & 0.21     \\ 
&ST  &  0  &  8.01 & 0.59 & 8.92 & 0.44 & 9.59 & 0.29    &  20.88 &  0.26 \\ 
&ST  &  5  &  7.66 & 0.64 & 8.31 & 0.56 & 9.01 & 0.46      &  21.33 &  0.26 \\ 
&ST  &  15 &  7.65 & 0.64 & 8.32 & 0.56 & 9.08 & 0.45    &  21.33 &  0.26 \\ 
&ST  &  60 &  7.71 & 0.63 & 8.32 & 0.55 & 8.97 & 0.45     &  21.18 &  0.26 \\ 
&STX &  h  &  7.76 & 0.63 & 8.32 & 0.56 & 9.17 & 0.44    &  21.10 &  0.26
\end{tabular}\end{center}\end{table}

{Table \ref{tab:CV} gives the prediction performance for each model and prediction lag. For comparison with the raw data measured in parts per billion (ppb), we exponentiate our model predictions (which were made based on the log-transformed data) and compute root mean square prediction errors (RMSPEs).}  As expected, the Car AB predictions are less accurate for all models, because they are usually made at longer spatial distances.  The static models (X and S) have lower correlation between predicted and observed than the spatiotemporal models, although the difference is less extreme for Car AB predictions. For these data the ST and STx models are similar for forecasting, although for small block sizes there are slight improvements in correlation by including the covariates in the covariance in the Car AB predictions, where exploiting long-range relationships explained by covariates may be useful. 

For the spatiotemporal models the correlations are considerably lower for the conditioning set that includes the nearest neighbors with $l=0$ versus conditioning sets with lag $l>0$.  The results are fairly similar for $l=\{5,15,60\}$ (and hence only shown for $l=h$ for STx) and so there does not seem to be sensitivity to the choice of lag apart from using a non-zero value. 
It is somewhat surprising that predictions are only slightly more accurate for larger block sizes than the raw one-second data.  For example, the model's highest one-hour ahead correlation is 0.31 for one-second data compared to 0.46 for one-minute data.  This is encouraging because it suggests that producing very high-resolution maps is feasible using these data.

Parameter estimates for the ST model are shown in Table \ref{tab:cor_est}. The spatial and temporal range estimates for the 15-second and one-minute block medians are generally longer than the estimates for the original one-second data. Therefore, it appears NO$_2$ is more stable after smoothing via block medians.  The range estimates vary considerably by the conditioning set lag $l$.  For $l=0$, and thus the closest neighbors included in the conditioning set, temporal (and spatial) range estimates are very small, indicating that only observations taken in the few minutes prior to the prediction are useful.  In contrast, the temporal range estimates are larger for $l>0$ and in some cases observations from the previous day remain correlated with the current observation. {This indicates that it is necessary to take the
conditioning set lag $l$ to be non-zero for estimation and prediction.}

\begin{table}
\caption{\textbf{Correlation parameter estimates.} Estimated spatial variance ratio ($\sigma^2/[\tau^2+\sigma^2]$), spatial ($\theta_x$) and temporal ($\theta_t$) range parameter estimates for one-second, fifteen-second, and one-minute block median datasets for the spatiotemporal model fit using the observations from the previous $[l,l+60]$ minutes in the conditioning set.}
\label{tab:cor_est}
\begin{center}\begin{tabular}{cc|rrr}
Block size & $l$ & Variance ratio & Spatial range (km) & Temporal range (hour)\\\hline
1 sec  &  0 & 1.00 & 0.95 & 0.19 \\
          &  5 & 0.63 & 4.82 & 3.83 \\
          & 15 & 0.54 & 3.95 & 9.53 \\
          & 60 & 0.77 & 1.38 & 2.32 \\
&\vspace{-6pt}\\
15 sec &  0 & 1.00 & 1.51 &  0.09\\
          &  5 & 0.56 & 8.01 & 20.64\\
          &  15 & 0.48 & 7.24 & 32.14\\
          &  60 & 0.52 & 2.11 &  4.30\\
&\vspace{-6pt}\\
1 min  &  0 & 0.92 & 3.52 &  0.23\\
          &  5 & 0.64 & 5.21 & 9.24\\
          &  15 & 0.57 & 5.43 & 28.72\\
          &  60 & 0.60 & 3.62 & 4.19\\
\end{tabular}\end{center}
\end{table}

\begin{figure}
 \centering
 \includegraphics[width=\textwidth]{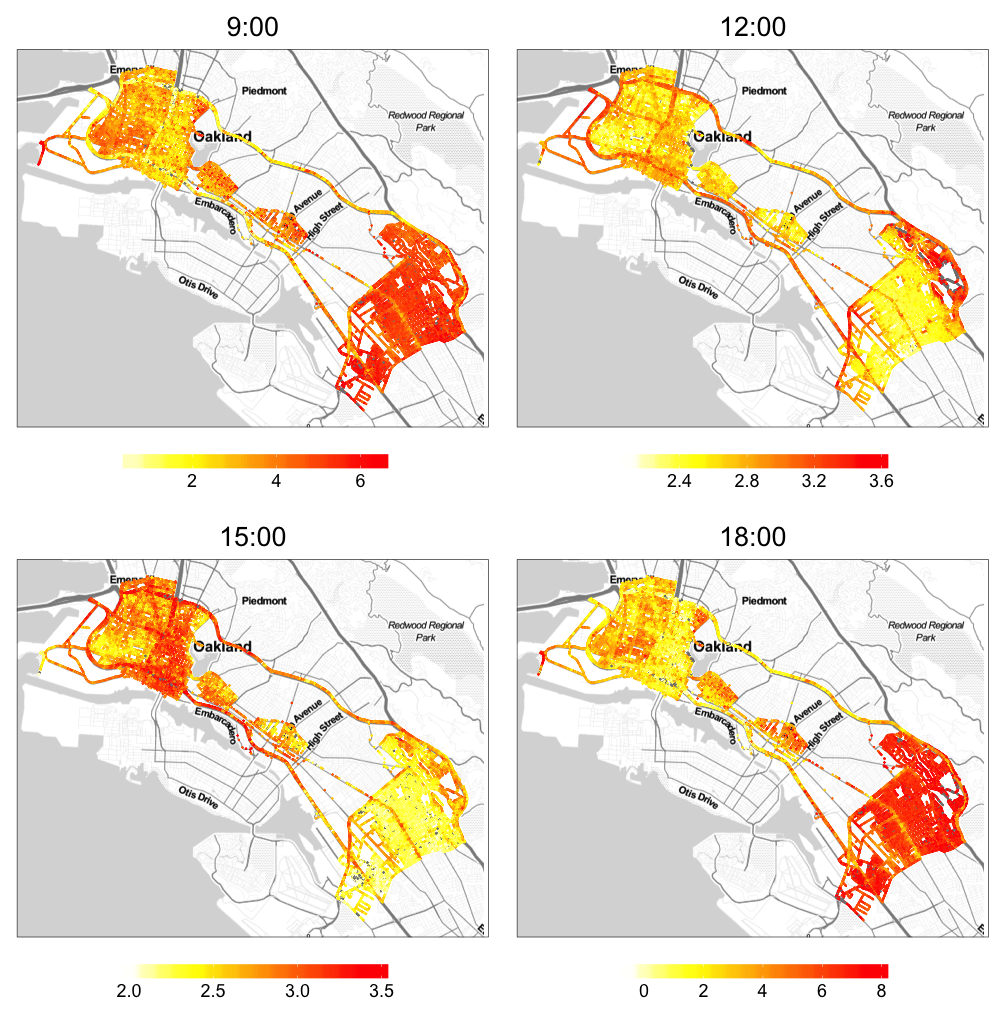}
 \caption{{\bf Estimated spatial diurnal pattern}: Maps of the estimated mean trend $\bX_t(\bs){\hat \bbeta}$ for several hour of the day for the ST model.   Note that because the scales vary so dramatically by hour the panels have different scales. Map tiles by \href{http://stamen.com}{Stamen Design}, under \href{http://creativecommons.org/licenses/by/3.0}{CC BY 3.0}. Data by \href{http://openstreetmap.org}{OpenStreetMap}, under \href{http://www.openstreetmap.org/copyright}{ODbL}.}
 \label{fig:mean}
 \end{figure}

Figure \ref{fig:mean} shows that the estimated mean trend $\bX_t(\bs){\hat \bbeta}$ varies considerably throughout the day. Southeast Oakland's Business District shows the largest diurnal fluctuation with high values during the rush hours of 9:00 and 18:00 and moderate levels midday.   Traffic patterns in Northwest Oakland are likely affected by Oakland City Center amenities such as the Oakland Convention Center and may be somewhat alleviated by access to multiple stops of the Bay Area Rapid Transit System. The diurnal fluctuations and rush-hour traffic patterns are also apparent in subregions of the NW, although not to as great of an extent. During midday (12:00 and 15:00) the high mean trends are along the freeways and the NE.  

\begin{figure}
\centering
\subfloat[]{\includegraphics[width=0.33\textwidth]{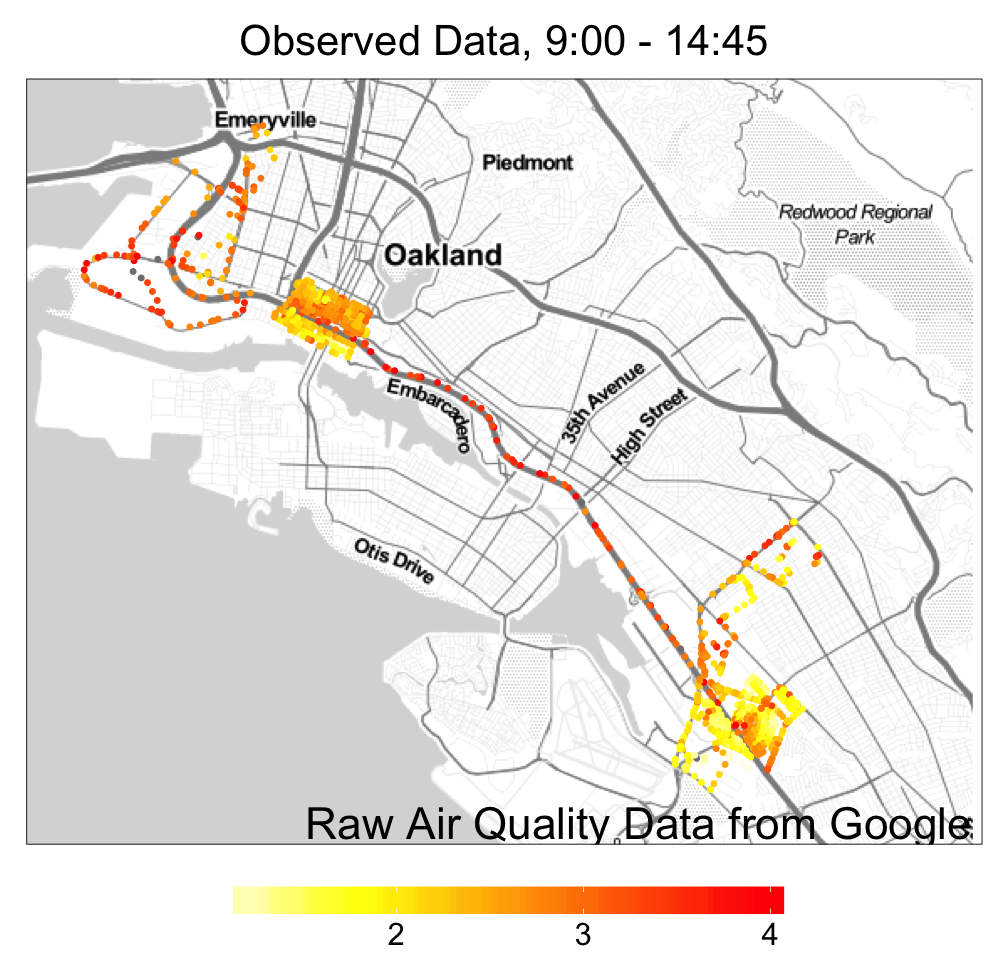}}
\subfloat[]{\includegraphics[width=0.33\textwidth]{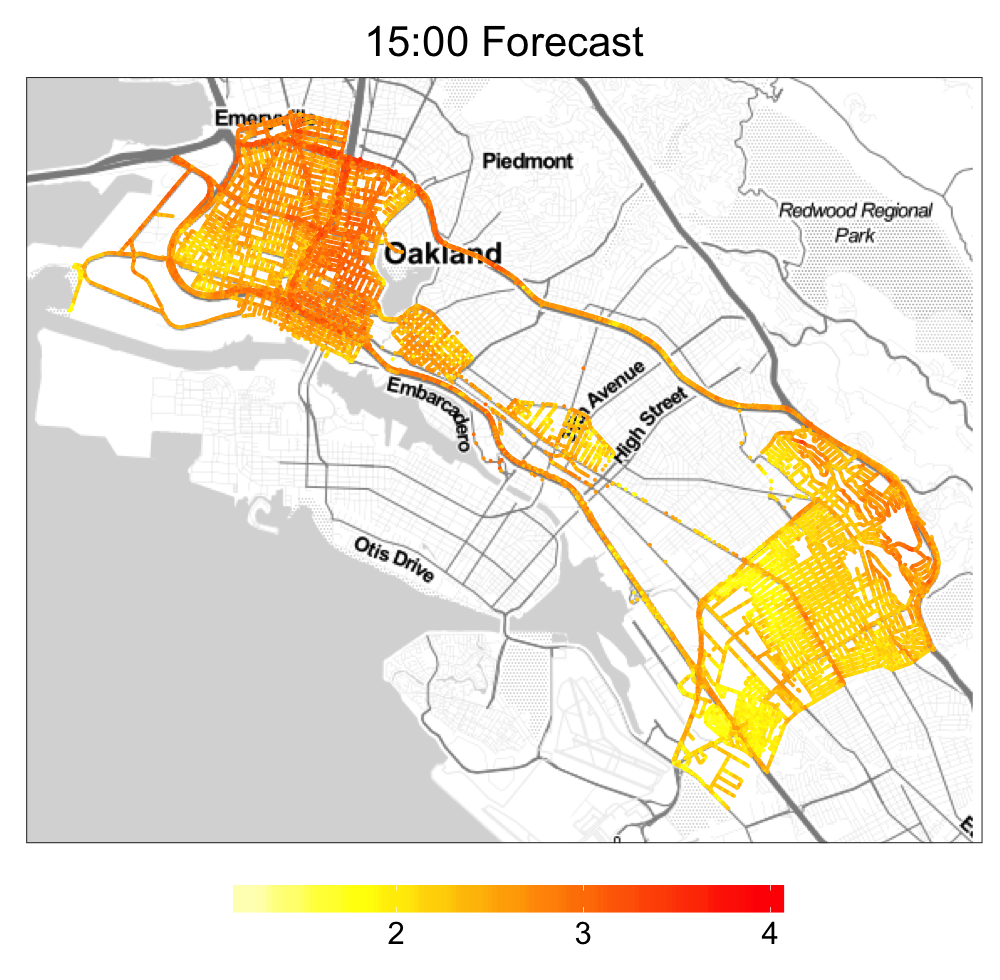}}
\subfloat[]{\includegraphics[width=0.33\textwidth]{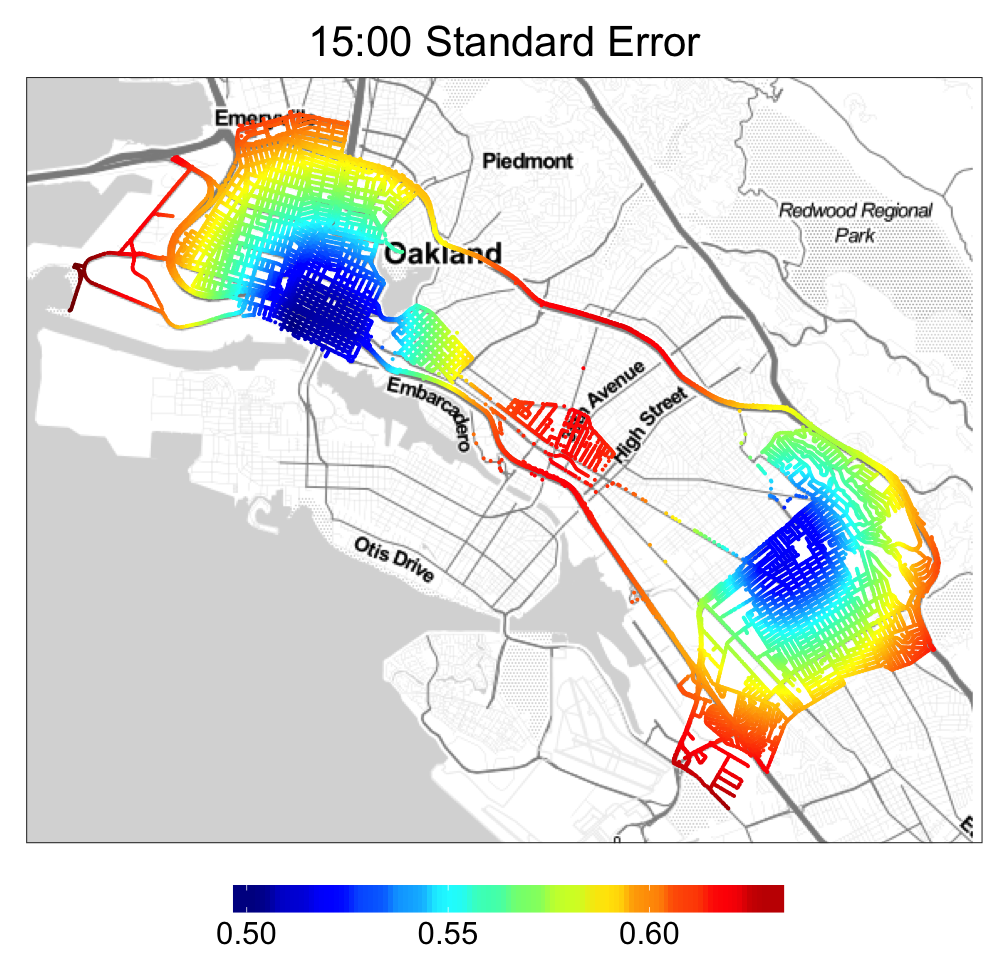}}
\caption{{\bf Real-time Forecasting Illustration}: (a) 15-second aggregated data from Cars A and B up to 14:45 on May 5, 2016. (b) 15-minute ahead forecasted log(NO$_2$) at 15:00 using the ST model. (c) Associated standard errors of log(NO$_2$) forecasts. Map tiles by \href{http://stamen.com}{Stamen Design}, under \href{http://creativecommons.org/licenses/by/3.0}{CC BY 3.0}. Data by \href{http://openstreetmap.org}{OpenStreetMap}, under \href{http://www.openstreetmap.org/copyright}{ODbL}.}
\label{fig:forecast}
\end{figure}

To illustrate how short-term predictions might look to a user, Figure \ref{fig:forecast} shows 15-minute ahead forecasted log(NO$_2$) at 15:00 using the data from 13:45 to 14:45 on May 5, 2016 using the ST model (b) with associated standard errors (c). The observed log(NO$_2$) obtained from Cars A and B earlier in the day (up to 14:45) are also shown in Figure \ref{fig:forecast}(a) for comparison. As expected, prediction standard errors are lowest where data has been obtained most recently from the two cars. 

\begin{figure}
\centering
\subfloat[]{\includegraphics[width=0.33\textwidth]{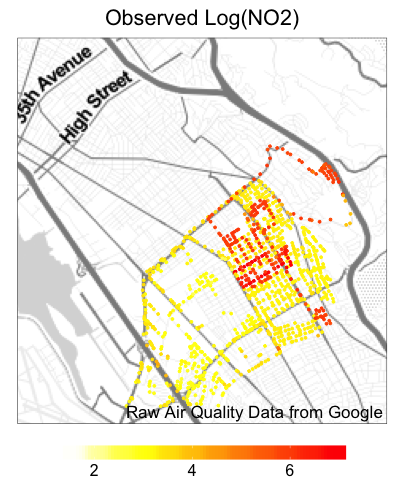}}
\subfloat[]{\includegraphics[width=0.33\textwidth]{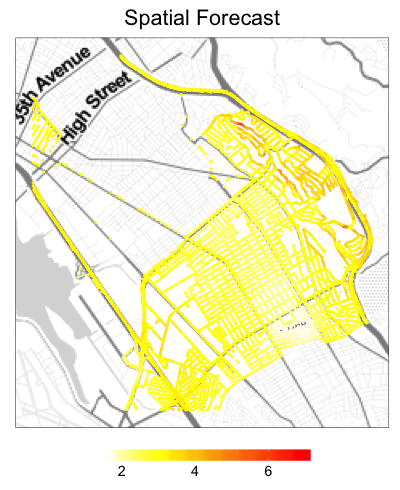}}
\subfloat[]{\includegraphics[width=0.33\textwidth]{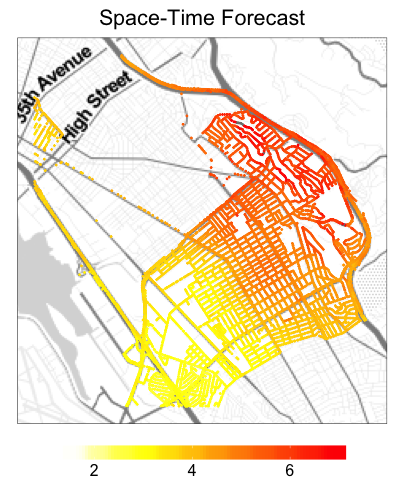}}

\subfloat[]{\includegraphics[width=0.33\textwidth]{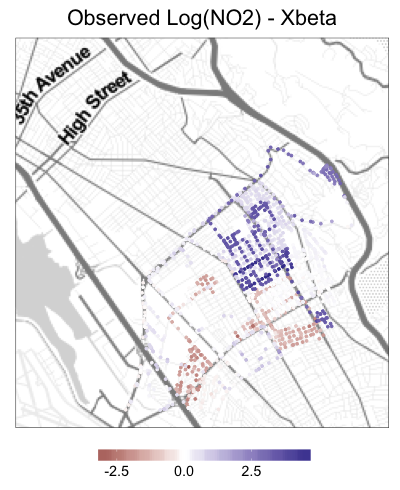}}
\subfloat[]{\includegraphics[width=0.33\textwidth]{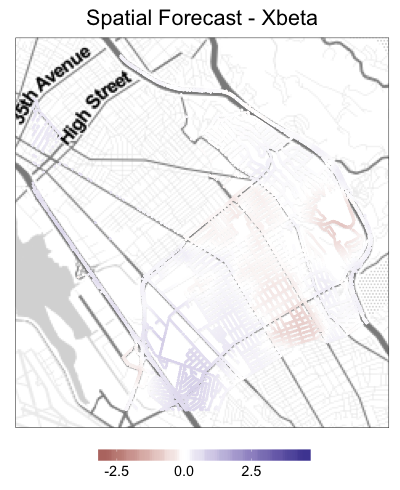}}
\subfloat[]{\includegraphics[width=0.33\textwidth]{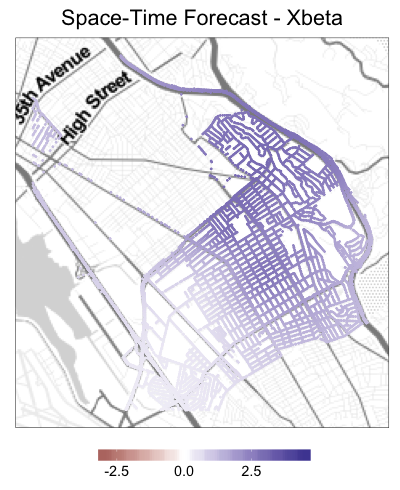}}
\caption{\textbf{Forecasting Model Comparison:} 15-minute ahead forecast of log(NO$_2$) for the southeast portion of the spatial domain at 14:00 using data up to 13:45 on September 11, 2015 (a) using the S (b) and ST (c) models. Observations (d) and forecasts (e), (f) with the estimated regression mean removed. Map tiles by \href{http://stamen.com}{Stamen Design}, under \href{http://creativecommons.org/licenses/by/3.0}{CC BY 3.0}. Data by \href{http://openstreetmap.org}{OpenStreetMap}, under \href{http://www.openstreetmap.org/copyright}{ODbL}.}
\label{fig:comp_forecast}
\end{figure}

Figure \ref{fig:comp_forecast} compares ST model 15-minute forecasts to S model forecasts to illustrate the flexibility of ST model to adapt to real-time observations with large deviations from the mean trend. Shown are observed log(NO$_2$) data of Oakland from 9:00 to 13:45 on September 11, 2015 (a) and the 15-minute ahead forecast at 14:00 from the S (b) and ST (c) models. The bottom row of Figure \ref{fig:comp_forecast} shows the deviation of the observed data and the 15min ahead forecasts from the estimated regression mean (i.e. $y_t(\bs) - x_t(\bs) \hat{\beta}$ or $\hat{y}_t(\bs) - x_t(\bs) \hat{\beta}$). The nearest observed data in time to 14:00 is substantially higher than the estimated mean over the region and this trend is reflected in the ST model forecasts. However, the S model forecasts do not exploit the real-time information from nearest neighbors in time and as such the S model is not able to provide realistic short-term forecasts of log(NO$_2$). This illustrates the power of the spatiotemporal dependence structure for real-time forecasting of the proposed model.

\subsection{Sliding window analysis}\label{s:window}

We envision the model being refitted periodically to adapt to evolving environmental, traffic and emissions patterns.  We refit the model using 15-second block median data in a sliding window of training data to study changes in parameter estimates and relative model performance.   For each week from 12/07/2015 to 05/13/2016, we use the data from the previous $w$ weeks to train the model and compute 15-minute ahead prediction mean squared error for that week as in Section \ref{s:compare}.  For the spatiotemporal models we use a one-hour lag in the conditioning set, so that $l=60$ in parameter estimation.   

\begin{figure}
\centering
\subfloat[]{\includegraphics[width=0.5\textwidth]{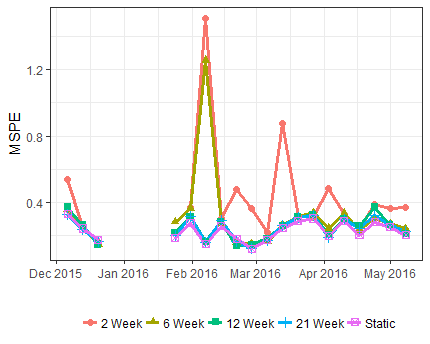}}

\subfloat[]{\includegraphics[width=0.5\textwidth]{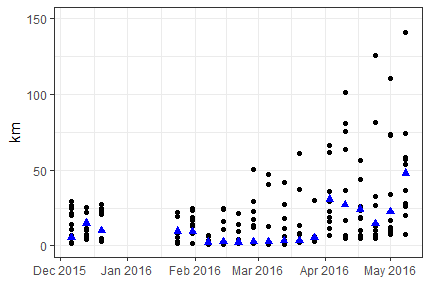}}
\subfloat[]{\includegraphics[width=0.5\textwidth]{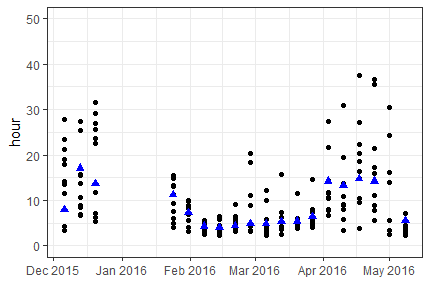}}
\caption{{\bf Sliding-window results}: \textbf{Mean square prediction error of log(NO$_2$)} (a) for the spatiotemporal (``ST'') model using sliding window with different sizes for training, and ST static model with parameter estimated from the first 21 weeks. Estimate (triangle) and bootstrap samples (circle) for the spatial (b) and temporal (c) range parameters assuming a 21-week window.}
\label{fig:window}
\end{figure}

The STX model has similar performance to the ST model, so we only present the results from the ST models. {Figure \ref{fig:window} plots the MSPEs of log(NO$_2$) using sliding window with sizes} $w = 2, 6, 12, 21$ weeks. {We compare the sliding-window results using MSPEs of log(NO$_2$) instead of MSPEs of the NO$_2$, because the latter results in large spikes that make it hard to see the differences. The sliding-window predictions are also compared with the prediction using static parameter values}, which are estimated using the first 21 weeks of training data from  07/14/2015 to 12/04/2015 and kept constant for the remaining period. While the magnitude of the errors varies considerable across weeks, the larger sliding window provides more stable and smaller MSPEs. The window size for training data should be large enough so that the cars have driven most of the area, since the parameters, especially the mean, can only be estimated reasonably well in this case. For our application, a sliding window of six weeks seems to be an appropriate size, as increasing the window size does not seem to improve prediction for most the weeks; however, caution should be taken when there are large gaps where we have no data, as this can result in poor parameter estimates and therefore, unreliable forecast such {as the week of February 15-19, 2016}. For the time period of our application, the sliding window approach performs similarly to the static approach, indicating stationarity in time; however, this may not hold for data collected over a longer time period, under which cases a sliding window approach will be more appropriate.  

Figure \ref{fig:window} plots the estimated spatial and temporal range parameters over time for the 21-week sliding window.  We assess the uncertainty in estimating the spatial and temporal range parameters using bootstrap samples. For each week, we randomly sample days from the previous 21 weeks with replacement to estimate the ST model parameters 15 times. The bootstrap estimates are plotted over time in Figure \ref{fig:window}. While the uncertainties are large, both spatial and temporal dependence are strong and change over time. 

\subsection{Experiments with different number of mobile and fixed-location monitors}\label{s:experiment}

If mobile devices are to be used to monitor a city's air pollution, natural questions are how many cars are needed to achieve the desired level of precision with respect to predictions and spatial estimation, and how does this mobile fleet perform compared to a network of fixed-location monitors in terms of prediction and spatial interpolation. Here, we conduct an experiment to compare the benefits of mobile versus fixed-location monitors. For different numbers of mobile and fixed-location monitors, we compute the expected mean square error for both spatial interpolation and short-term forecasting.    

\begin{figure}
\centering
\includegraphics[width=0.65\textwidth]{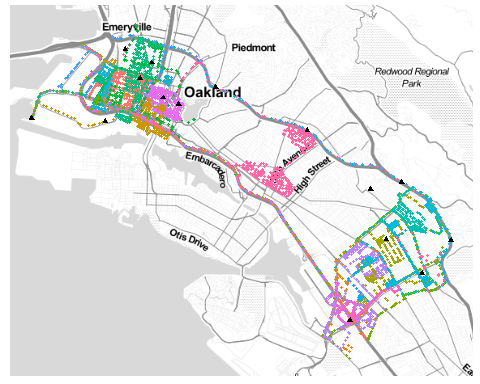}
\caption{{\bf Network design}: Locations of fixed-location monitors (triangle) and a sample of mobile monitor routes (each route is a different color). Map tiles by \href{http://stamen.com}{Stamen Design}, under \href{http://creativecommons.org/licenses/by/3.0}{CC BY 3.0}. Data by \href{http://openstreetmap.org}{OpenStreetMap}, under \href{http://www.openstreetmap.org/copyright}{ODbL}.}
\label{fig:network}
\end{figure}

To emulate real Google Street View data, we sample $c$ car routes from the days when the car is in service from 13:00-16:00. This time period is limited in terms of practicality for fully characterizing air quality trends, but is selected for this limited demonstration because the majority of drive time on any given day covers this interval.  We treat the $c$ car routes as if they were observed simultaneously. For each route, we record the location and the time of the day, collecting data every 15 seconds. This is repeated 30 times to assess the uncertainty, due to randomness in the routes, in prediction performance, which crucially depends on the amount of area covered by each route. To create fixed-location monitor data, we randomly select $m$ locations as monitoring sites; while fixed point monitors usually evaluate NO$_2$ over 1 hour averages, these sites are assumed to have sampling frequency of 15 seconds in order to match the sampling frequency of the mobile monitors for comparison. Figure \ref{fig:network} shows the deployment of mobile and fixed-location monitors.

Let $\bY_{T^\ast} = \left\lbrace Y_{T^\ast}(\bs): \bs\in\mathcal{S}^\ast\right\rbrace$ be the prediction set, where $T^\ast$ is the prediction time and $\mathcal{S}^\ast = \left\lbrace \bs_1^\ast, \dots,  \bs_n^\ast \right\rbrace$ are $n=2000$ chosen locations throughout Oakland. Since Kriging is unbiased, the expected MSE is simply the predictive variance, which depends only on the locations of the observations and the model parameters and is thus available without having to simulate data.  We use the spatiotemporal covariance parameters estimated from the data analyzed in Section \ref{s:compare} (Table \ref{tab:cor_est}) to compute the average prediction variance,
\begin{equation}
\text{MSPE}_{T^\ast}^{(\mathcal{J})} = \frac{1}{n} \sum_{i=1}^n \text{Var}[Y_{T^\ast}(\bs_i^\ast) \mid \bY^{(\mathcal{J})}] = \sigma^2 + \tau^2 - \frac{1}{n} \sum_{i=1}^n \bSigma_{iT^\ast\mathcal{J}} (\bSigma_{\mathcal{J}}+\tau^2 I)^{-1}\bSigma_{iT^\ast\mathcal{J}}^T, 
\end{equation}
where $\bSigma_{iT^\ast\mathcal{J}} = \mbox{Cov}[Y_{T^\ast}(\bs_i^\ast), \bY^{(\mathcal{J})}]$, $\bSigma_{\mathcal{J}} = \mbox{Cov}(\bY^{(\mathcal{J})})$, and $\mathcal{J}\in\{\mbox{mobile},\mbox{stationary}\}$ indicates whether the conditioning set $\bY^{(\mathcal{J})}$ is from mobile or fixed-location monitors. 

\begin{figure}
\centering
\includegraphics[width=0.65\textwidth]{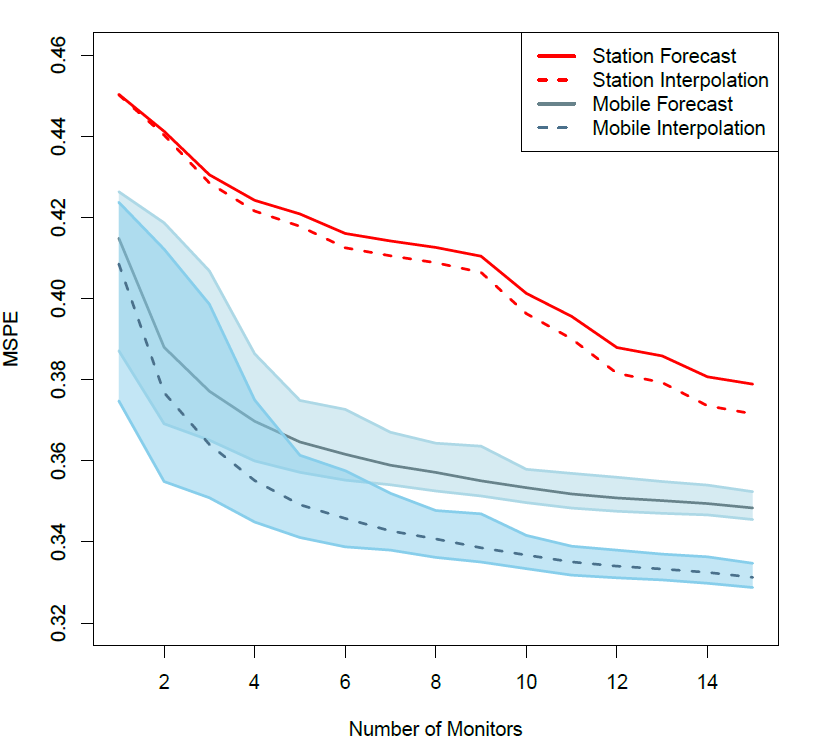}
\caption{{\bf Evaluation of network design}: Mean squared prediction error (MSPE) \textbf{in log(NO$_2$)} by the number of mobile versus fixed-location monitors for short-term forecasting and spatial interpolation. The 95$\%$ credible intervals of MSPE for mobile monitors are represented by polygons.}
\label{fig:exp}
\end{figure}

Two types of prediction are compared: 30-minute-ahead forecasting and spatial interpolation.  The 30-minute-ahead forecast uses the data from 13:00-15:30 to map NO$_2$ over Oakland for $T^\ast$ = 16:00; for spatial interpolation we use data from 13:00-15:30 to map NO$_2$ at time $T^\ast$ = 14:15.  Results are shown in Figure \ref{fig:exp}.  The MSPE for forecast and interpolation are similar because the temporal range estimated from the data using the two-step procedure indicates a long time dependence of about 17 hours.  As the number of monitors (either mobile and stationary) increases the MSPE decreases and levels off around 5 monitors.  The 95$\%$ confidence interval of MSPE for uncertainty due to route randomness is wider when the number of mobile monitors is small; however, as more mobile monitors are deployed, a larger spatial domain is covered by the routes, consequently reducing the uncertainty.  It appears that using more than five mobile monitors reduces the MSPE by only a small margin and therefore are not useful for these settings.  The reduction in prediction variance, as indicated by the slope in Figure \ref{fig:exp}, is much faster for mobile monitors than monitors at fixed locations.  Under this specific setting, using only 3-4 mobile monitors gives comparable prediction performance as 15 monitors at fixed but random locations. 

\section{Conclusions}\label{s:conclusions}

Methodology to provide real-time air pollution maps as well as short-term air quality forecasts on a fine-resolution temporal and spatial scale can dramatically improve understanding of local environments.  This paper contributes to the emerging field of mobile air pollution monitoring by providing a template for processing and modeling data with complex measurement and scale considerations using a unique source of highly detailed data with spatial and temporal complexities.  
We addressed the practical considerations of temporal aggregation and spatial neighborhood scheme for local approximation to optimize short-term forecasting. Our approach has forecast skill, outperforming competing methods. {The proposed two-step procedure utilizing the Vecchia likelihood approximation scales linearly with the sample size and can therefore be implemented on very large data sets. Increasing the size of the neighboring set will increase computational burden, but we found that for the data we considered, taking the neighbor size to be $m = 30$ for two mobile sensors gave similar parameter estimates as $m = 60$, and therefore a larger $m$ appeared to be unnecessary.  Similarly, the window size also affects computational time, but we found six weeks to be sufficient for our application. If a larger window size is used, increasing the number of cores for computing the approximated likelihood in parallel will also reduce overall computational time. Lastly, the temporal ordering we considered for the Vecchia approximation is intuitive given the limited spatial coverage at a single time point, but a more sophisticated space-time ordering scheme could easily be implemented if necessary for a larger number of concurrently deployed mobile sensors.}

{Forecasting air pollution using the data currently available from the Google Street View experiment does have its limitations. In our analyses, we utilize solely the data available from the Google Street View vehicles, which have limited spatial and temporal coverage.
The mobile measurements are also inherently noisy and highly variable due to local and unpredictable phenomena. Therefore, the forecast air quality maps should be further calibrated with other data sources, such as near-road monitors and station monitors. A possible future extension of the methodology is to jointly model data from mobile and station monitors, in which case it would be crucial to account for differences in the bias and variance of each data source.}

The experiment conducted to examine the relative efficiency of a fleet of mobile monitors versus a network of monitors at fixed locations finds that mobile monitors can provide comparable estimation and prediction when resources are limited. {However, distinctions between the network of monitors at fixed locations and stationary monitors designed and operated by local governments should be made clear, because the two differ substantially in measurement technology, sampling frequency, and data quality. In the experiment we make the critical assumption that the mobile and fixed-location monitors provided data of the same quality, but this assumption is currently not necessarily true in practice. An experiment to make informed decisions about the future deployment of stationary versus mobile monitors in practice would need to incorporate the relative quality of the data associated with each monitoring system. }

{As data production and real-time availability continue to be driven by the ongoing development and improvement of mobile measurement technology, the modeling framework developed has important real-world implications in better understanding local environments. For example, the number of available mobile sensors may dramatically increase as data collection functionality on individuals' smartwatches improves. Extending the proposed methodology to account for more mobile sensors should be conceptually straightforward, but determining the optimal neighboring and lag structure would require careful study.}

\section*{Acknowledgements}
We would like to thank Karin Tuxen-Bettman and Joshua Apte for their help in obtaining the Google Street View Air Quality data and Alison Eyth and Madeleine Strum of EPA for emissions expertise.  Google Street View data can be requested from Google and EDF by interested users at 
\href{https://www.edf.org/airqualitymaps/oakland/download-oakland-air-pollution-data}{\nolinkurl{https://www.edf.org/airqualitymaps/oakland/download-oakland-air-}\\
             {\nolinkurl{pollution-data}}
             }.
The views, opinions, and observations expressed in this article are of the authors and are not necessarily that of the EPA nor the NSF. This work was partially supported by the National Institutes of Health (R01ES027892), the National Institutes of Environmental Health Sciences (K99ES029523) and the National Science Foundation (DMS-1638521 and DMS-1654083).

\begin{singlespace}
	\bibliographystyle{rss}
	\bibliography{refs}
\end{singlespace}

\section*{Appendix A.1: Simulation results}\label{s:appendix}
In this section we present a short simulation study to investigate the relationship between the lag used to fit the model and misspecification of the covariance structure.  Time series data are generated as 
$$Y_t = 0.9 Y_{t-1} + \theta Z_{t-1} + \theta Z_{t-2} + Z_t$$
where $Z_t\iid\mbox{Normal}(0,1)$. Data are generated using two values of $\theta$: $\theta=0$ gives an AR(1) model and $\theta=0.9$ give an ARMA(1,2) model.  We fit the model assuming an AR(1) structure using a simple linear regression of $Y_t$ onto $Y_{t-l}$ for fitting lag $l$. Denoting ${\hat b}_l$ as the estimated slope, lag-$h$ predictions of $Y_t$ given $Y_{t-h}$ are then made as ${\hat Y}_t = {\hat b}_l^{h/l} Y_{t-h}$. Each simulated dataset consists of 10,000 observations for fitting the model and 10,000 additional observations to evaluate prediction mean squared error.   We repeat this experiment with $\theta \in \{0,0.9\}$ and $h\in\{1,5,10,20\}$.  For each scenario we generate 1000 datasets and Table \ref{tab:app_sim} presents the prediction MSE for $l\in\{2,5,10,20\}$ relative to the MSE for $l=1$.  

\begin{table}
\caption{\textbf{Simulation study results.} Lag-$h$ predication mean squared error (MSE) for data generated from either an AR(1) or ARMA(1,2) model.  Parameters are estimated assuming an AR(1) model using simple linear regression with lag $l$.  MSE is presented relative to the MSE of the fit with $l=1$.}
\label{tab:app_sim}
\centering
\begin{tabular}{ll|cccc}
&& \multicolumn{4}{c}{$l$}\\
Model & $h$ & 2 & 5 & 10 & 20\\\hline
    AR(1) &  1 & 1.000 & 1.000 & 1.000 & 1.001 \\ 
    AR(1) &  5 & 1.000 & 1.000 & 1.000 & 1.002 \\ 
    AR(1) & 10 & 1.000 & 1.000 & 1.000 & 1.002 \\ 
    AR(1) & 20 & 1.000 & 1.000 & 1.000 & 1.001 \\ 
ARMA(1,2) &  1 & 1.007 & 1.038 & 1.053 & 1.065 \\ 
ARMA(1,2) &  5 & 0.957 & 0.940 & 0.942 & 0.946 \\ 
ARMA(1,2) & 10 & 0.924 & 0.889 & 0.888 & 0.890 \\ 
ARMA(1,2) & 20 & 0.907 & 0.876 & 0.874 & 0.874 \\ 
\end{tabular}\end{table}

When the model is specified correctly, i.e., the data are generated from an AR(1) process with $\theta=0$, then all methods perform similarly for all prediction lags.  However, when the model is misspecified, i.e., the data are generated from an ARMA(1,2) model, then fitting with lag $l=h$ gives the best predictions.  In these most extreme cases, for lag $h=1$ predictions, fitting with lag $l=20$ gives 6.5\% higher MSE than fitting with $l=1$; for lag $h=20$ predictions, fitting with lag $l=1$ gives 12.6\% higher MSE than fitting with lag $l=20$.  Therefore, this simple simulation study shows that fitting a model that excludes the most highly-correlated observations from model fitting can improve long-range prediction.  

\end{document}